\documentclass[preprint,12pt]{elsarticle}

\usepackage{graphics}
\usepackage{graphicx}
\usepackage{epsfig}
\usepackage{amssymb}
\usepackage{epstopdf}
\def\eq#1{{Eq.~(\ref{#1})}}
\def\fig#1{{Fig.~\ref{#1}}}

\journal{Nuclear Physics A}

\begin{document}

\begin{frontmatter}



\author{Javier L. Albacete}
\ead{javier.lopez-albacete@cea.fr}
\address{Institut de Physique Th\'eorique, CEA/Saclay, 91191 Gif-sur-Yvette cedex, France}
\author{Cyrille Marquet}
\ead{cyrille.marquet@cern.ch}
\address{Physics Department, Theory Unit, CERN, 1211 Gen\`eve 23, Switzerland}

\title{Single and double inclusive particle production in d+Au collisions at RHIC, leading twist and beyond}


\begin{abstract}
We discuss the evidence for the presence of QCD saturation effects in the data collected in d+Au collisions at RHIC. In particular we focus our analysis on forward hadron yields and azimuthal correlations. Approaches alternative to the CGC description of these two observables are discussed in parallel.
\end{abstract}


\end{frontmatter}


\section{Introduction}
\label{intro}

The large amount of experimental data collected at RHIC during the last decade over a wide kinematical range allows to explore novel QCD effects. Indeed RHIC measurements present a number of features which are well described within the Color Glass Condensate effective theory of QCD at high energies (see e.g.
\cite{Gelis:2010nm,Weigert:2005us} and references therein). The main physical ingredient in the CGC is the inclusion of unitarity effects through the proper consideration of both non-linear recombination effects in the quantum evolution of hadronic wave functions and multiple scatterings at the level of particle production. Such effects are expected to be relevant when partons with a small enough energy fraction $x$ are probed in the nuclear (or, in full generality, hadronic) wave function.

The CGC formalism itself is only reliable at small $x$, in that kinematic regime gluon densities are large and gluon self-interactions become highly probable, thus taming, or saturating, further growth of the gluon occupation numbers. While the need for unitarity effects comprised in the CGC is, at a theoretical level, clear, the real challenge from a phenomenological point of view is to assess to what extent they are present in available data. Such is a difficult task, since different physical mechanisms concur in data, and also because the limit of asymptotically high energy in which the CGC formalism is developed may not be realized in current experiments.

Notwithstanding these difficulties, we argue below that RHIC data
\cite{Arsene:2004ux,Adams:2006uz,Braidot:2010zh} offer compelling evidence for the presence of saturation effects. Such claim is based on the successful simultaneous description of the suppression of particle production \cite{Albacete:2010bs} and azimuthal correlations \cite{Albacete:2010pg} at forward rapidities in d+Au collisions compared to p+p collisions, using the most up-to-date theoretical tools available in the CGC approach. We focus on forward particle production for the following reason: such processes are sensitive only to high-momentum partons inside one of the colliding hadron, which appears dilute, while mainly small-momentum partons inside the other dense hadron contribute to the scattering. Since the high$-x$ part of a proton wave function is well understood in perturbative QCD, forward particle production in high-energy d+Au (or generically p+A) collisions is ideal to investigate the small$-x$ part of the nucleus wave function.

In the case of single-inclusive hadron production $pA\!\to\!hX$, denoting $p_{\perp}$ and $y$ the transverse momentum and rapidity of the final state particles, the partons that can contribute to the cross section have a fraction of longitudinal momentum bounded from below, by $x_p$ (for partons from the proton wave function)
and $x_A$ (for partons from the nucleus wave function) given by
\begin{equation}
x_p=x_F\ ,\hspace{0.5cm}x_A=x_F\ e^{-2y}\ ,
\label{kin1}
\end{equation}
where we introduced the Feynman variable $x_F=|p_{\perp}|e^{y}/\sqrt{s_{NN}}$ with $\sqrt{s_{NN}}$ the collision energy per nucleon. With $\sqrt{s_{NN}}\gg|p_{\perp}|$ and forward rapidities $y\!>\!0,$ the process features $x_p\!\lesssim\!1$ and $x_A\!\ll\!1,$ meaning that the scattering involves on the proton side, quantum fluctuations well understood in QCD, and on the nucleus side, quantum fluctuations whose non-linear QCD dynamics can be studied.

In the case of double-inclusive hadron production $pA\!\to\!h_1h_2X$, denoting $p_{1\perp},$ $p_{2\perp}$ and $y_1,$ $y_2$ the transverse momenta and rapidities of the final-state particles, the Feynman variables are $x_i=|p_{i\perp}|e^{y_i}/\sqrt{s_{NN}}$ and $x_p$ and $x_A$ read
\begin{equation}
x_p=x_1+x_2\ ,\hspace{0.5cm}
x_A=x_1\ e^{-2y_1}+x_2\ e^{-2y_2}\ .
\label{kin2}
\end{equation}
Therefore, only the forward-forward case is sensitive to values of $x$ as small as in the single particle production case: $x_p\!\lesssim\!1$ and $x_A\!\ll\!1$. The central-forward measurement does not probe such
kinematics: moving one particle forward increases significantly the value of $x_p$ compared to the
central-central case (for which $x_p=x_A=|p_{\perp}|/\sqrt{s_{NN}}$), but decreases $x_A$ only marginally. For this reason we shall refer to these two situations as mid-rapidity ones. The approximations made in CGC calculations apply best with forward-rapidity observables, at RHIC mid-rapidity ones are contaminated too much by large-$x$ physics to be treated at a quantitative level.

Saturation-based approaches were the only ones to correctly predict the suppression of forward particle production \cite{Kharzeev:2003wz,Albacete:2003iq} and the azimuthal decorrelation of forward hadron pairs
\cite{Marquet:2007vb}. In the following, we also comment on alternative mechanisms that successfully describe mid-rapidity particle production in d+Au collisions. We note that, while some of these approaches are also able to describe, a posteriori, the suppression of forward yields, we are not aware of any formalism that can also describe the azimuthal decorrelation of forward hadron pairs.

\section{Nuclear modifications at mid-rapidity}

In relativistic heavy-ion collisions, nuclear effects on single particle production are typically evaluated in terms of ratios of particle yields called nuclear modification factors:
\begin{equation}
R^h_{pA}=\frac{dN^{pA\to hX}/dyd^2p_\perp}{N_{coll}\ dN^{pp\to hX}/dyd^2p_\perp}\ ,
\end{equation} 
where $N_{coll}$ is the number of nucleon-nucleon collisions in the p+A collision. If high-energy nuclear reactions were a mere incoherent superposition of nucleon-nucleon collisions, then the observed $R_{pA}$ would be equal to unity. However, RHIC measurements in d+Au collisions (or peripheral Au+Au collisions)
\cite{Arsene:2004ux,Adams:2006uz} exhibit two opposite regimes: at mid rapidities the nuclear modification factors exhibit an enhancement in particle production at intermediate momenta $|p_\perp|\sim 2\div 4$ GeV. In turn, a suppression at smaller momenta is observed. However, at forward rapidities the Cronin enhancement disappears, turning into an almost homogeneous suppression.

\begin{figure}[t]
\begin{center}
\includegraphics[height=3.9cm]{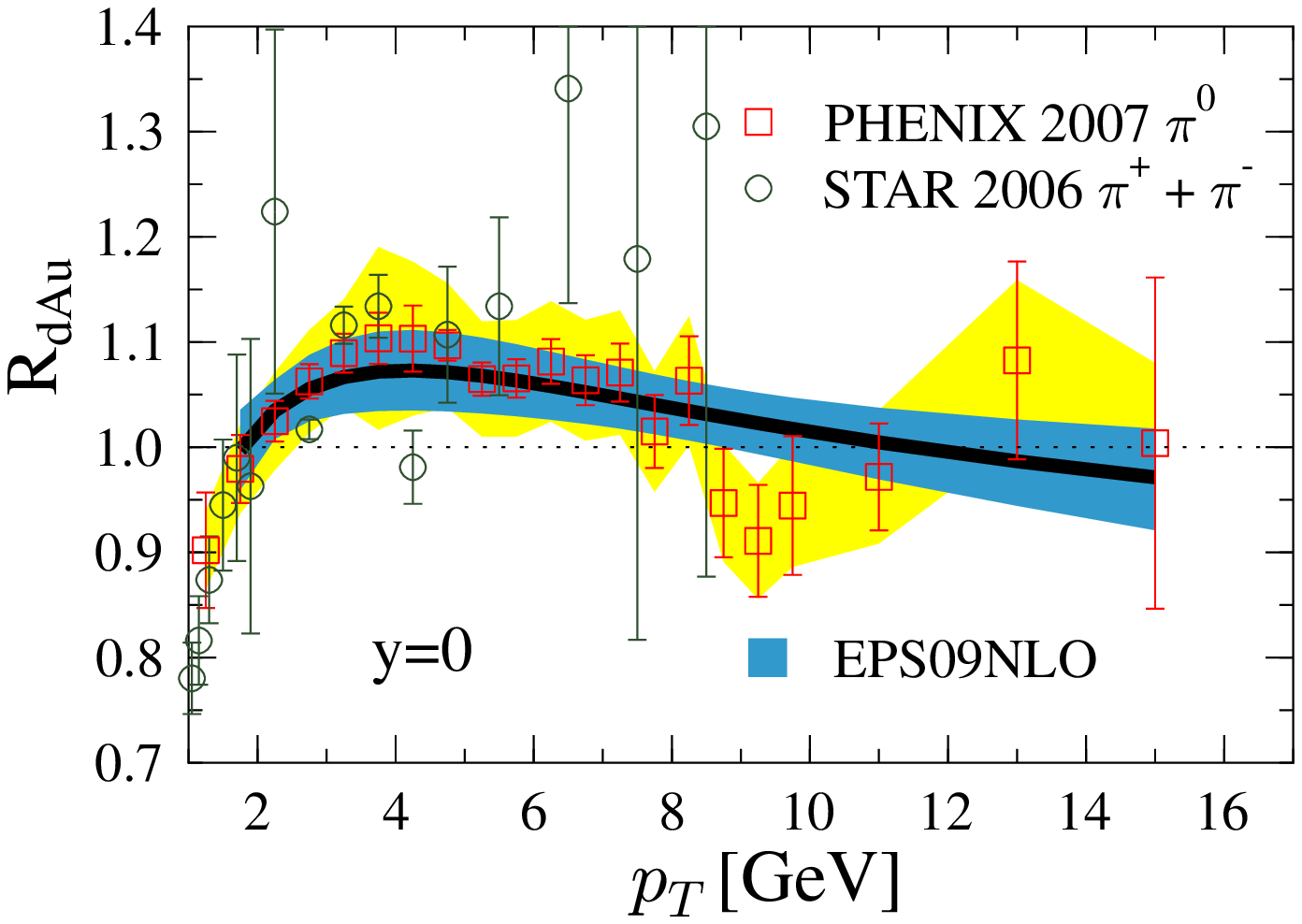}
\includegraphics[height=3.8cm]{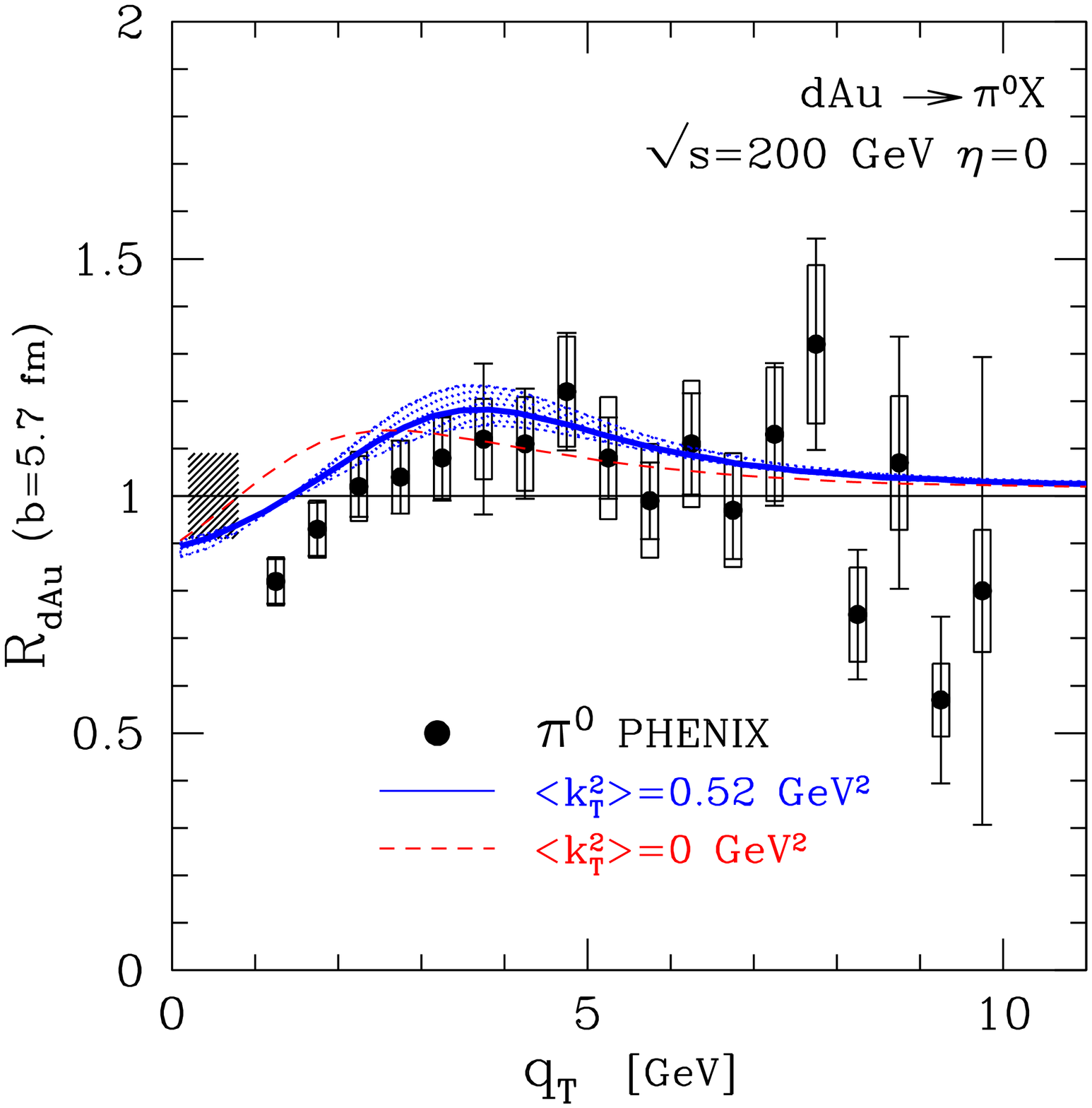}
\includegraphics[height=4.0cm]{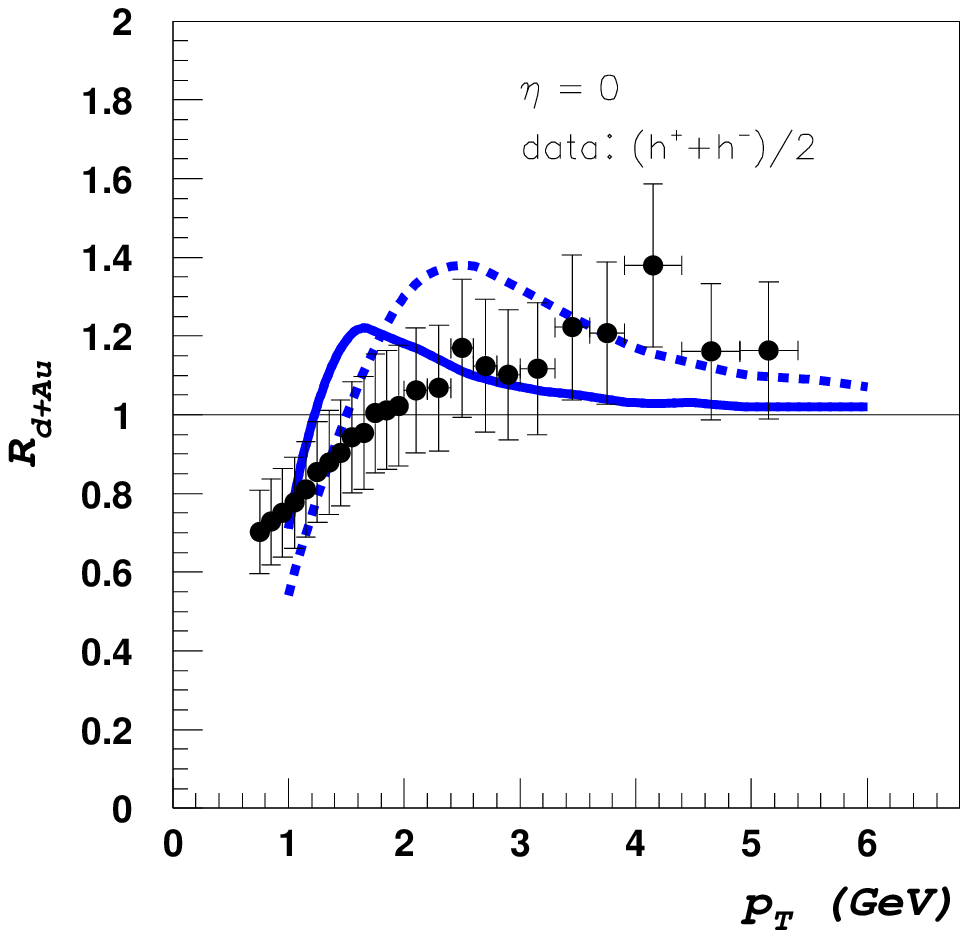}
\end{center}
\caption{RHIC experimental results for $R_{dAu}^{\pi^0}$ at $\eta=0$ compared with different calculations.
From left to right: comparison with EPS09 parametrization \cite{Eskola:2009uj}, a Glauber-Eikonal calculation \cite{Accardi:2003jh}, and the CGC approach of \cite{Kharzeev:2004yx}.}
\label{eta0}
\end{figure}

Single-hadron production data at mid-rapidity have been successfully analyzed through different formalisms and techniques. Below we sketch an incomplete, but representative list of the variety in the theory spectrum: 
\begin{itemize}
\item {\bf Leading-twist perturbation theory}. The assumption is that standard collinear factorization holds in nuclear reactions, meaning that highly-virtual partons in nuclei behave independently as they do in protons. For each parton species $i$ the nuclear parton distribution functions are taken to be proportional to that of a proton: $f_i^A(x,Q^2)=R_i^A(x,Q^2)\,f_i^N(x,Q^2)$. The proportionality factors $R_i^A(x,Q^2)$ are fitted, in part, to available d+Au data and in some cases, as in the EPS parametrization \cite{Eskola:2009uj}, evolved according to DGLAP evolution. The resulting data description is displayed in \fig{eta0}a.

\item {\bf Glauber-eikonal multiple scatterings}. This approach takes into account power corrections to the leading-twist approximation. It relies on a resummation of incoherent multiple scatterings. Typically, this results in a momentum broadening of the scattered parton which is responsible for the Cronin enhancement and, due to unitarity constraints, to a depletion of particle production at small transverse momenta, in agreement with the qualitative features of the data as can be seen in \fig{eta0}b. Performing the complete resummation including energy-momentum conservation is a challenging task. Sometimes, a detour of the strict calculation is taken by resorting to unintegrated parton distributions which include information about the intrinsic transverse momentum of the partons $k_0$, and the average transverse momentum gained during the interaction with the nucleus $\Delta k$:
$f_i^A(x,Q^2)\rightarrow F_i^A(x,Q^2,<\!k_0^2\!>\!+\!<\!\Delta k^2\!>(\sqrt{s},b,p_\perp)$. While the intrinsic
$k_0$ is adjusted in p+p collisions, the gained transverse momentum is let to depend on the collision energy, centrality and $p_\perp$ of the detected hadron.

\item {\bf Color Glass Condensate}. The CGC approach resums all power corrections which are dominant in the high-energy/small-$x$ limit. It relies on two main assumptions: the scattering process is fully coherent and the propagation of the leading parton through the nucleus is eikonal, i.e. the momentum transfered during the collision is only transverse. Then, non-linearities or saturation effects can be taken into account either at the semiclassical level or, more accurately, including the quantum corrections embodied in the JIMWLK evolution equation. The work presented in \fig{eta0}c relies on a combination of both, with quantum corrections modeled according to analytical estimates \cite{Kharzeev:2004yx}. Note that in the CGC framework, the saturation scale $Q_s$ which characterizes the onset of non-linear effects in the nuclear wave function, also determines both the intrinsic transverse momentum and the amount of it gained during the collision. In the small-$x$ limit, it is not possible to distinguish saturation from multiple scatterings, such a separation would be frame dependent. Both become important when a large gluon density is reached, and including one without the other is not consistent.

\end{itemize}

Simply by looking at Fig. 1, one concludes that the three different approaches above offer a comparably good description of the data, so it is difficult to extract any clean conclusion about the physical origin of the Cronin enhancement. This is probably due to the kinematic region probed in these measurements. For a hadron momentum of $|p_\perp|\sim 2$ GeV, one is sensitive to $(x_p\sim)x_A\sim 0.01\div0.1$. In this region different physical mechanisms concur, so neither of the physical assumptions underlying the approaches above is comletely fulfilled. Indeed, the coherence length, estimated to be $l_c\sim 1/(2m_N\,x_A)\sim 1 \div 10$ fm is, on average, smaller than the nuclear length, so the fully coherent treatment of the collision implicit in the CGC is not completely justified, and large-$x$ corrections are expected to be relevant. Moreover, both the coherent or incoherent treatments of the collision in \cite{Accardi:2003jh,Kharzeev:2004yx} need to invoke the presence of an intrinsic scale, presumably of non-perturbative origin, of the order of 1 Gev in order to reproduce the data, whose dynamical origin is not evident at all.

\begin{figure}[t]
\begin{center}
\includegraphics[height=6.2cm]{fwd-cent-1.eps}
\hfill
\includegraphics[height=6.3cm]{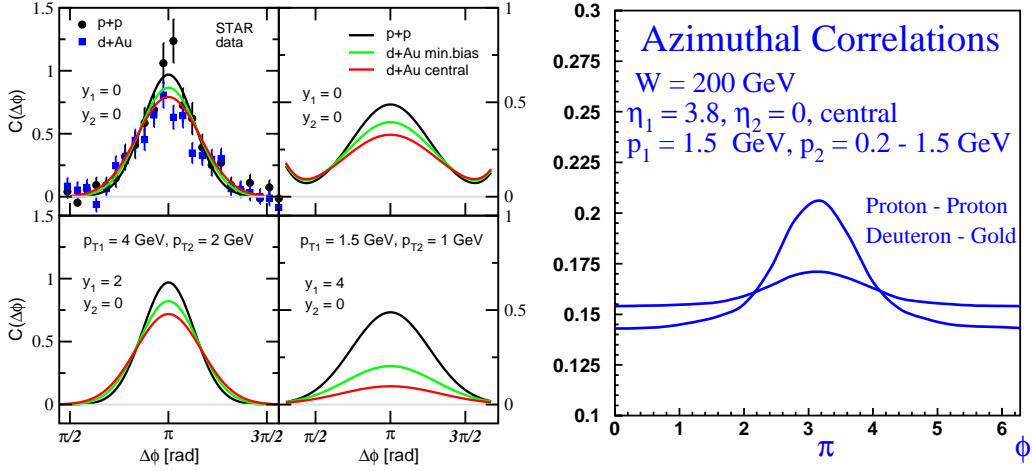}
\end{center}
\caption{The coincidence probability at mid-rapidity as a function of $\Delta\phi$. RHIC data show that in the
central-central case the away-side peak is similar in d+Au and p+p collisions. In the central-forward case, the Glauber-eikonal (left plot from \cite{Qiu:2004da}) and CGC (right plot from \cite{Kharzeev:2004bw}) calculations predict that the away-side peak is suppressed in d+Au compared to p+p collisions, in agreement with later data.}
\label{fwd-cent}
\end{figure}

More insight is gained with di-hadron correlations. In particular, let us focus on the $\Delta\phi$ dependence of the double-inclusive hadron spectrum, where $\Delta\phi$ is the difference between the azimuthal angles of the measured particles $h_1$ and $h_2$. Nuclear effects on di-hadron correlations are typically evaluated in terms of the coincidence probability to, given a trigger particle in a certain momentum range, produce an associated particle in another momentum range. In a p+p or p+A collision, the coincidence probability is given by $CP(\Delta\phi)=N_{pair}(\Delta\phi)/N_{trig}$ with
\begin{equation}
N_{pair}(\Delta\phi)=\int\limits_{y_i,|p_{i\perp}|}\frac{dN^{pA\to h_1 h_2 X}}{d^3p_1 d^3p_2}\ ,\quad
N_{trig}=\int\limits_{y,\ p_\perp}\frac{dN^{pA\to hX}}{d^3p}\ .
\label{kinint}
\end{equation}
First measurements were performed at RHIC at mid-rapidity by the PHENIX and STAR collaborations
\cite{Adams:2006uz,Adler:2006hi}. In the central-central case, the coincidence probability features a near-side peak around $\Delta\phi=0,$ when both measured particles belong to the same mini-jet, and an away-side peak around $\Delta\phi=\pi,$ corresponding to hadrons produced back-to-back. In the central-forward case, there is naturally no near-side peak.

\begin{figure}[t]
\begin{center}
\includegraphics[height=6cm]{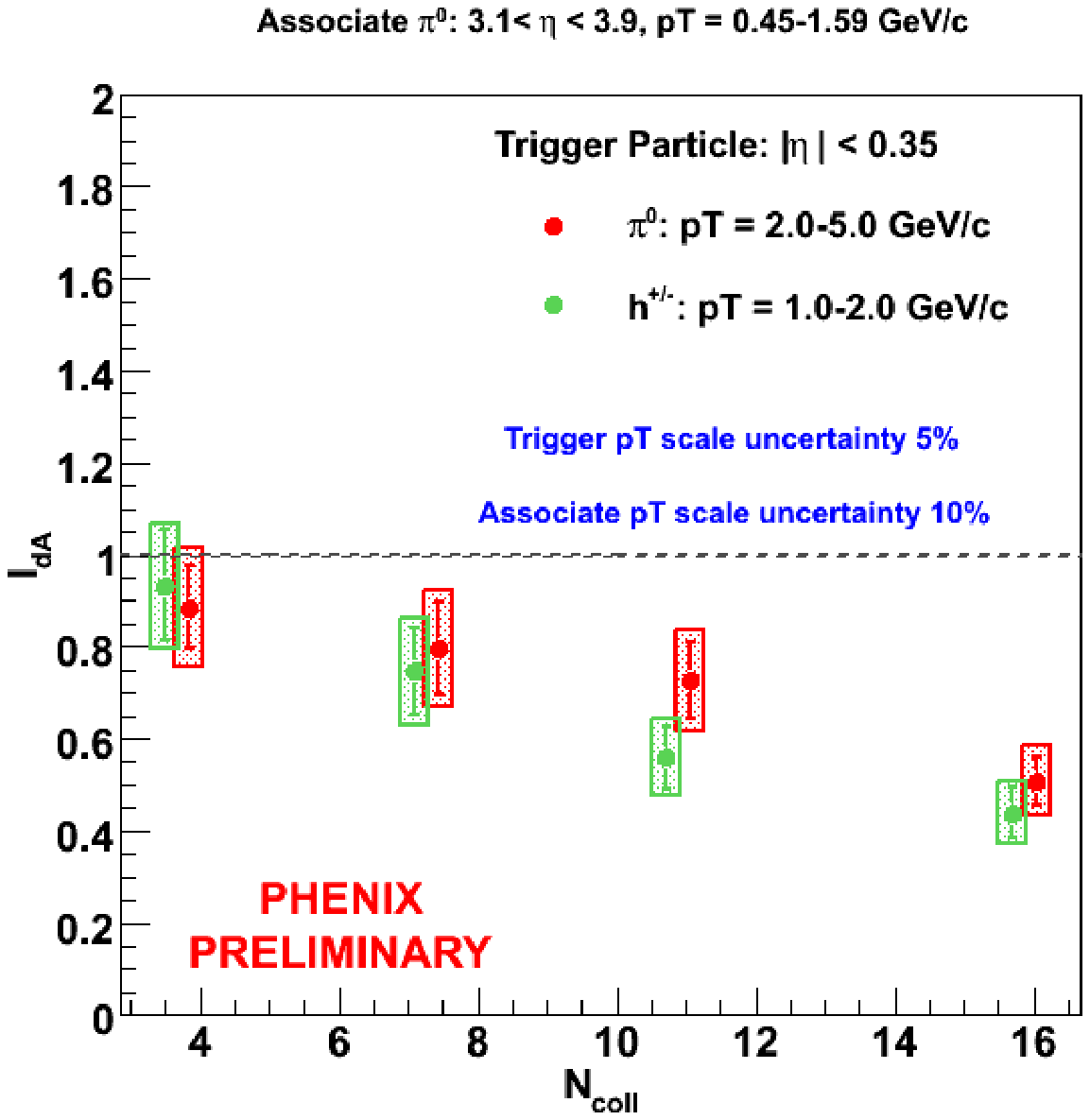}
\end{center}
\caption{Central-forward preliminary $I_{dAu}$ data as a function of centrality \cite{Meredith:2009fp}. In central collisions, the integral of the coincidence probabilty is about half that in p+p collisions, reflecting the depletion of the away-side peak. Collinear factorization cannot reproduce this behavior, while both Glauber-eikonal and CGC calculations predicted it.}
\label{IdA}
\end{figure}

Either in p+p or d+Au collisions, the sizeable width of the away-side peak cannot be described within the leading-twist collinear factorization framework. This indicates that, while it may not be obvious in single particle production, power corrections are important when $|p_\perp|\sim 2$ GeV. At such low transverse momenta, collinear factorization does not provide a global picture of particle production at RHIC, even at mid-rapidity. On the contrary, both the Glauber multiple scattering \cite{Qiu:2004da} and CGC
\cite{Kharzeev:2004bw} approaches can qualitatively describe the data, including the depletion of the away-side peak in d+Au collisions when going from central-central to central-forward production. This is illustrated in \fig{fwd-cent}. Such a depletion does not occur in p+p collisions, it is due to nuclear-enhanced power corrections, and therefore the p+A to p+p ratio of the integrated coincidence probabilites
\begin{equation}
I_{pA}=\frac{\int d\phi\ CP_{pA}(\Delta\phi)}{\int d\phi\ CP_{pp}(\Delta\phi)}
\end{equation}
is below unity. In \fig{IdA}, recent PHENIX data on $I_{dAu}$ are displayed as a function of centrality. At the moment, since $x_A$ is not that small, it is not clear whether the mechanism for this suppression is due to saturation effects rather than incoherent multiple scatterings.

\section{Moving forward}

As outlined in the introduction, data collected in the deuteron fragmentation region offer a cleaner opportunity to explore saturation effects. Let us first explain our implementation of the CGC framework. The CGC is equipped with a set of non-linear renormalization group equations to describe the evolution of hadronic wave functions towards small-$x$. In the large-$N_c$ limit, they reduce to the BK equation
\cite{Balitsky:1995ub,Kovchegov:1999yj}. The recent determination of running coupling corrections to the original leading-log equations \cite{Balitsky:2006wa,Kovchegov:2006vj} has proven an essential step in promoting the BK equation to a phenomenological tool. Indeed, the running coupling BK equation (rcBK) has been employed to successfully describe inclusive structure functions in e+p scattering \cite{Albacete:2009fh} and also the energy and multiplicity dependence of total hadron multiplicities in Au+Au collisions at RHIC
\cite{Albacete:2007sm}.

The rcBK equation reads
\begin{eqnarray}
  \frac{\mathcal{N}(r,Y)}{\partial\ln(1/x)}=\int d^2{\bf r_1}\
  K^{{\rm run}}({\bf r},{\bf r_1},{\bf r_2}) \left[\mathcal{N}(r_1,Y)+\mathcal{N}(r_2,Y)
\right.\nonumber\\\left.-\mathcal{N}(r,Y)-\mathcal{N}(r_1,Y)\,\mathcal{N}(r_2,Y)\right]\ ,
\label{bk1}
\end{eqnarray}
where ${\bf r_2}={\bf r}-{\bf r_1}$ (we use the notation $v\equiv |{\bf v}|$ for
two-dimensional vectors in \eq{bk1} and \eq{kbal}). $\mathcal{N}(r,Y)$ is the dipole scattering amplitude in the fundamental representation, with $Y=\ln(x_0/x)$ the rapidity and $r$ the dipole transverse size. It turns out that the evolution kernel
\begin{equation}
  K^{{\rm run}}({\bf r},{\bf r_1},{\bf r_2})=\frac{N_c\,\alpha_s(r^2)}{2\pi^2}
  \left[\frac{1}{r_1^2}\left(\frac{\alpha_s(r_1^2)}{\alpha_s(r_2^2)}-1\right)+
    \frac{r^2}{r_1^2\,r_2^2}+\frac{1}{r_2^2}\left(\frac{\alpha_s(r_2^2)}{\alpha_s(r_1^2)}-1\right) \right]
\label{kbal}
\end{equation}
proposed in \cite{Balitsky:2006wa} minimizes the role of higher order corrections, making it better suited for phenomenological applications. Detailed discussions about other prescriptions proposed to define the running coupling kernel, and about the numerical method to solve the rcBK equation can be found in
\cite{Albacete:2007yr}.

\eq{bk1} needs to be suplemented with initial conditions, which we choose to be of the McLerran-Venugopalan type:
\begin{equation}
\mathcal{N}(r,Y\!=\!0)=
1-\exp\left[-\frac{r^2\,\bar{Q}_{s0}^2}{4}\,\ln\left(\frac{1}{\Lambda\,r}+e\right)\right]\ ,
\end{equation}
where $\Lambda=0.241$ GeV. This introduces two free parameters: the value $x_0$ where the evolution starts and the initial saturation scale felt by quarks $\bar{Q}_{s0}$.

\subsection{Nuclear modification factors}

According to Ref.~\cite{Dumitru:2005gt}, the differential cross section for forward hadron production in
p+A collisions is given by 
\begin{eqnarray}
\frac{dN^{pA\to hX}}{dy\,d^2p_\perp}=K\sum_{q}\int_{x_F}^1\,\frac{dz}{z^2}\
\left[x_1f_{q\,/\,p}(\tilde{x}_p,p_\perp^2)\ F\left(\tilde{x}_A,\frac{p_\perp}{z}\right)\
D_{h\,/\,q}(z,p_\perp^2)\right.\nonumber\\ +\left. x_1f_{g\,/\,p}(\tilde{x}_p,p_\perp^2)\
\tilde{F}\left(\tilde{x}_A,\frac{p_\perp}{z}\right)\,D_{h\,/\,g}(z,p_\perp^2)\right]
\label{hyb}\ ,
\end{eqnarray}
where the unintegrated gluon distributions $F$ and $\tilde{F}$ are related to the dipole scattering amplitude through Fourier transformations: 
\begin{eqnarray}
F(x,k)=\int \frac{d^2{\bf r}}{(2\pi)^2}\
e^{-i{\bf k}\cdot{\bf r}}\left[1-\mathcal{N}(r,Y\!=\!\ln(x_0/x))\right]\ ,
\label{ugdfund}\\
\tilde{F}(x,k)=\int \frac{d^2{\bf r}}{(2\pi)^2}\
e^{-i{\bf k}\cdot{\bf r}}\left[1-\mathcal{N}(r,Y\!=\!\ln(x_0/x))\right]^2\ .
\label{ugdadj}
\end{eqnarray}
In principle $\tilde{F}$ is the Fourier transform of the dipole scattering amplitude in the adjoint representation \cite{Kovner:2001vi,Kovchegov:2001sc,Marquet:2004xa}, we have used to large-$N_c$ limit in
\eq{ugdadj}. In \eq{hyb}, $f_{i/p}$ and $D_{h/i}$ refer to the parton distribution function (pdf) of the incoming proton and to the final-state hadron fragmentation function respectively. Here we will use the CTEQ6 NLO pdf's \cite{Pumplin:2002vw} and the DSS NLO fragmentation functions
\cite{deFlorian:2007aj,deFlorian:2007hc}. We have assumed that the factorization and fragmentation scales are both equal to the transverse momentum of the produced hadron. Note that the projectile in our calculations is actually a deuteron, and we obtain the neutron pdf from the proton one assuming isospin symmetry.

For light hadron production discussed here, the difference between the rapidity and pseudo-rapidity $\eta$ of the produced hadron can be neglected, yielding the following kinematics: $\tilde{x}_p=x_F/z$ and
$\tilde{x}_A=(x_F/z)\exp{(-2y)}$ with
$x_F=\sqrt{m_h^2+p_\perp^2}/\sqrt{s_{NN}}\ \exp{(\eta)}\approx|p_\perp|/\sqrt{s_{NN}}\ \exp{(y)}$ introduced before. Due to parton fragmentation, the values of $\tilde{x}$'s actually probed are generically higher than $x_p$ and $x_A$ defined in \eq{kin1}.

With this set up we reach a very good description \cite{Albacete:2010bs} of the forward negatively charged hadron and neutral pion yields measured by the BRAHMS \cite{Arsene:2004ux} and STAR \cite{Adams:2006uz} collaborations respectively, in p+p and minimum bias d+Au collisions. The parameters of the fit are
$x_0=0.025\div 0.005$ (0.0075) and $\bar{Q}_{s0}^2=0.5\div0.4$ (0.2) GeV$^2$ for the initial nucleus (proton) wave function. Moreover, no $K$-factors are needed except for the most forward pions $\eta\!>\!3$, where we find that $K=0.3$ (0.4) is needed to describe the data. 

\begin{figure}[t]
\begin{center}
\includegraphics[height=4.5cm]{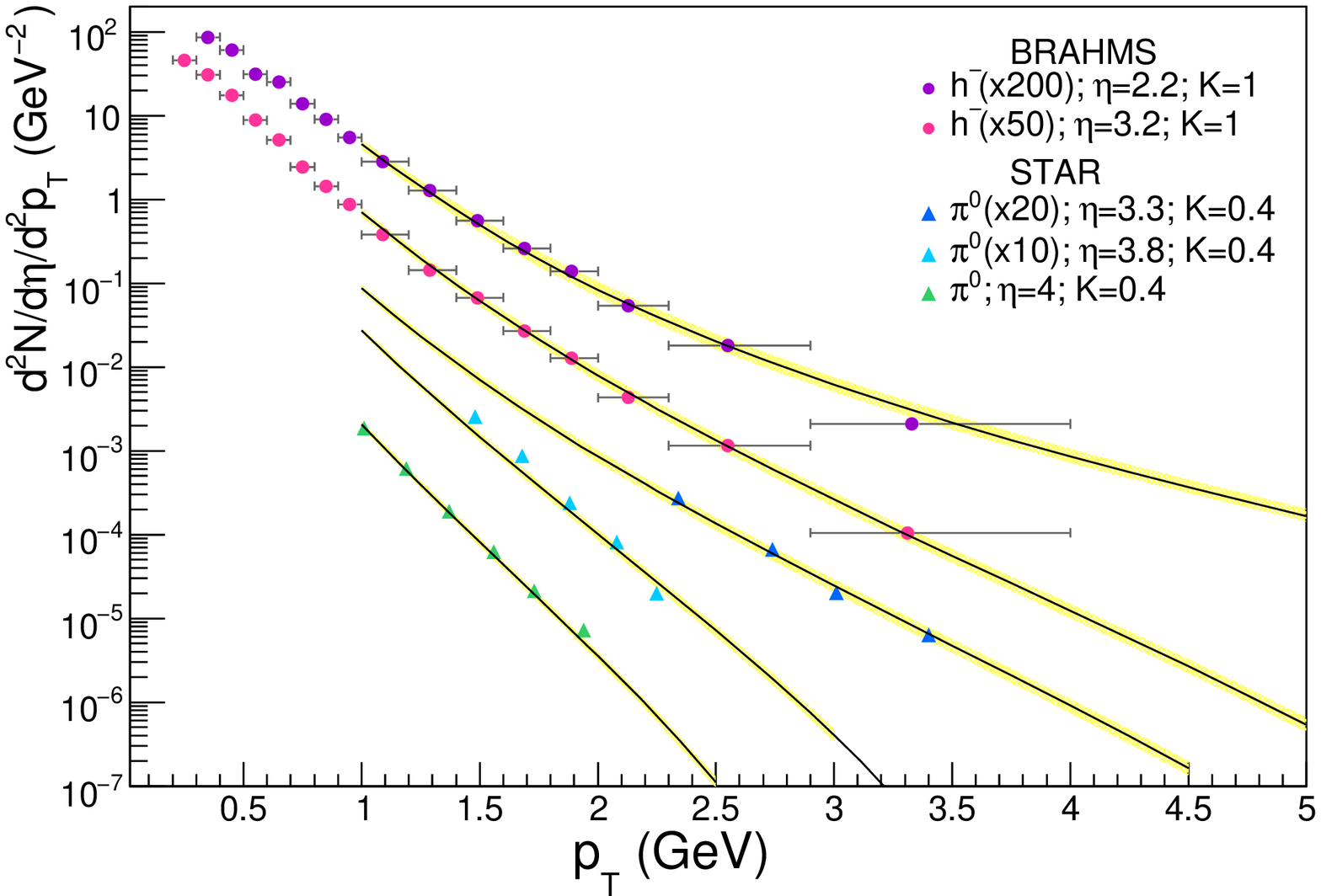}
\hfill
\includegraphics[height=4.5cm]{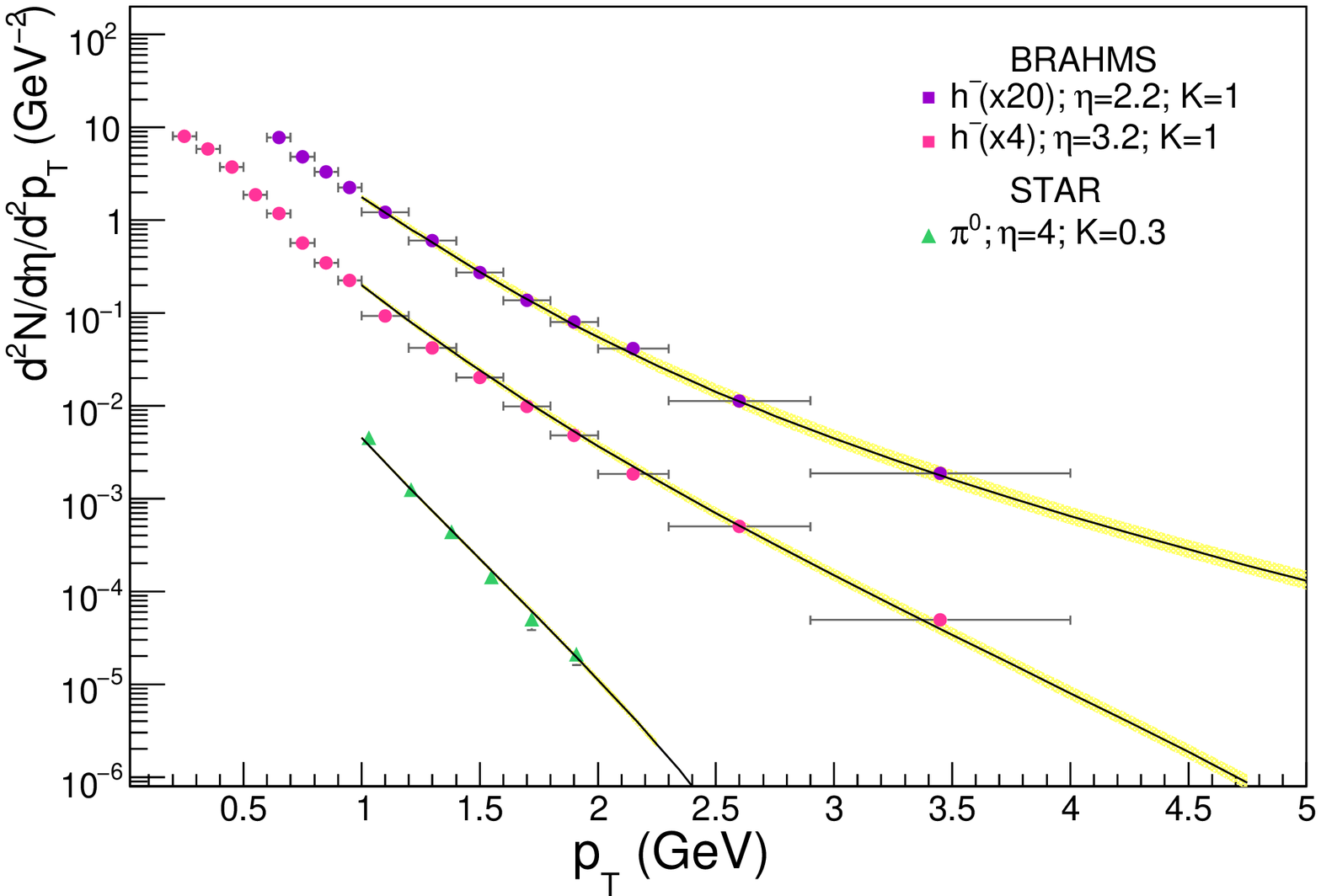}
\end{center}
\caption{Negatively charged hadrons and neutral pions at forward rapidities at RHIC in p+p (left) and minimum bias d+Au (right) collisions, compared to our calculation \cite{Albacete:2010bs}.}
\label{forward}
\end{figure}

\begin{figure}[t]
\begin{center}
\includegraphics[height=5cm]{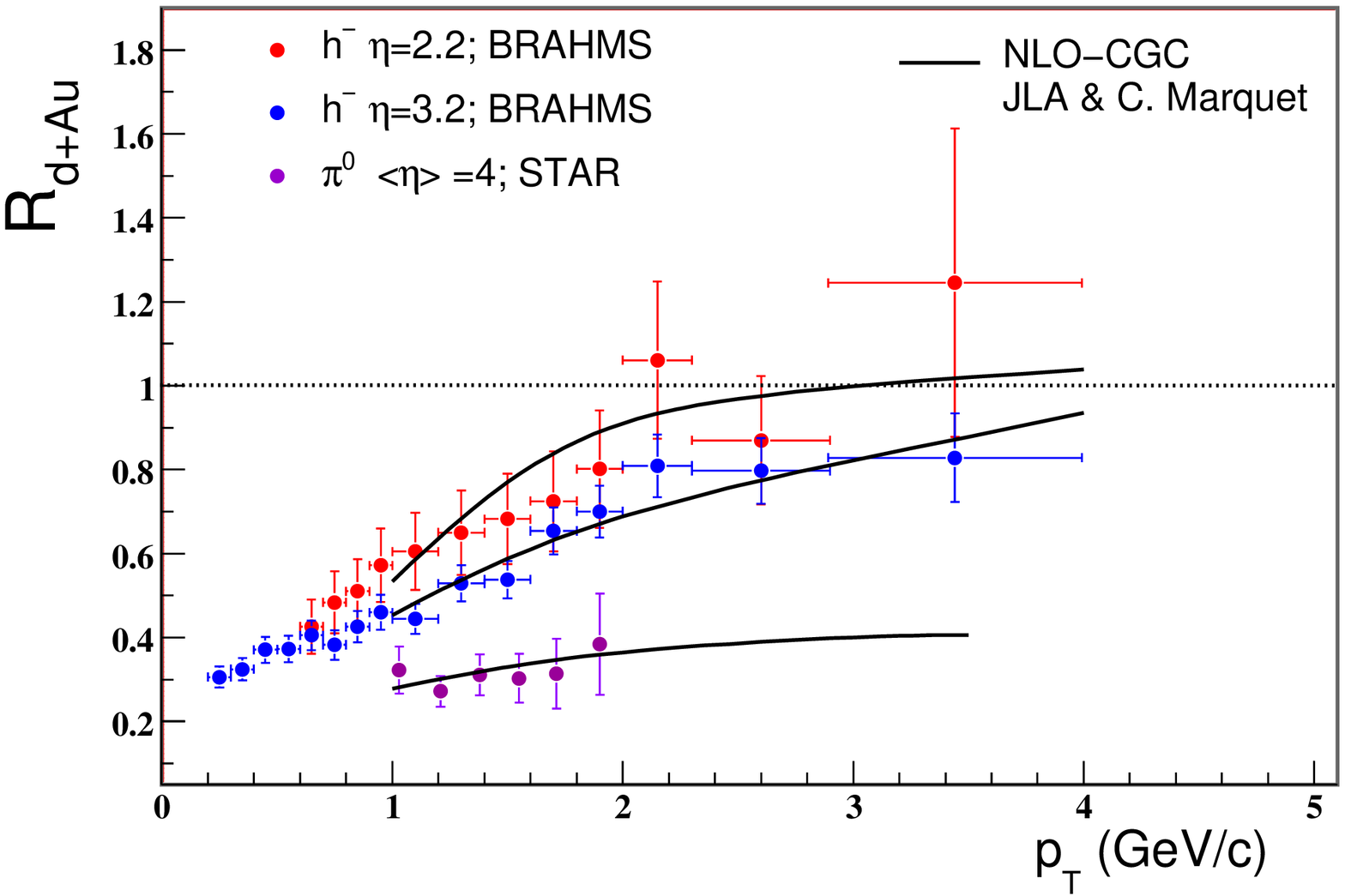}
\hfill
\includegraphics[height=5cm]{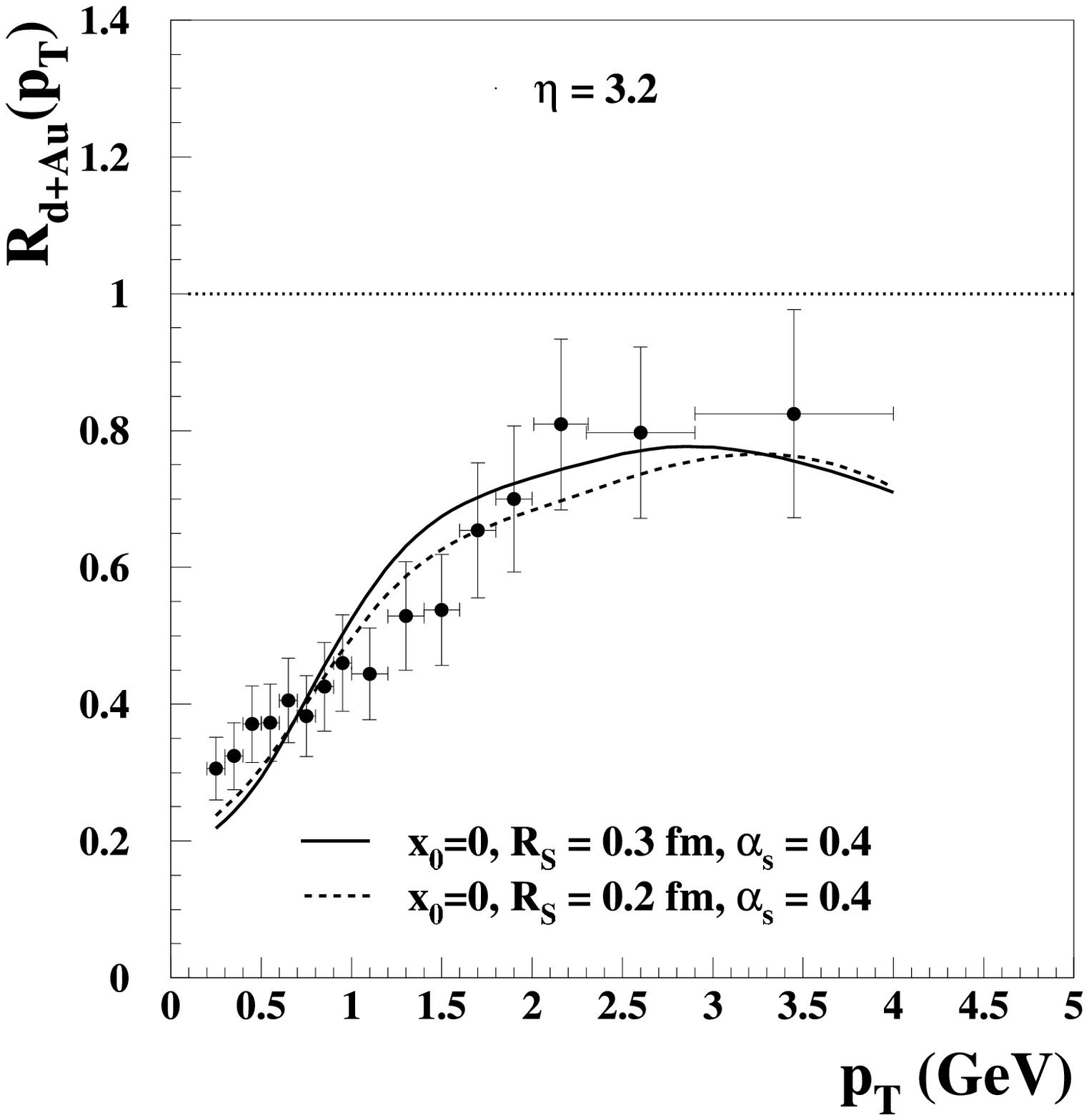}
\caption{Nuclear modification factors at forward rapidities in minimum bias d+Au collisions in CGC
\cite{Albacete:2010bs} (left) and energy-loss \cite{Kopeliovich:2005ym} (right) calculations. }
\label{ratios}
\end{center}
\end{figure}

Our results are shown in \fig{forward}. By simply taking the ratios of the corresponding spectra, we get a very good description of the nuclear modification factors at forward rapidities, and this is shown in \fig{ratios}a. It should be noted that we use the same normalization as experimentalists do in their analysis of minimum bias d+Au collisions: $N_{coll}=7.2$. Physically, the observed suppression is due to the relative enhancement of non-linear terms in the small-$x$ evolution of the nuclear wave function with respect to that of a proton.

After the data confirmed the CGC expectations, it has been argued that the observed suppression of particle production at forward rapidities is not an effect associated to the small values of $x_A$ probed in the nuclear wave function, but rather to energy-momentum conservation corrections relevant for $x_F\to 1$
\cite{Kopeliovich:2005ym,Frankfurt:2007rn}. Such corrections are not present in the CGC, built upon the eikonal approximation (this may explain why a $K$-factor is needed to describe the suppression of very forward pions). The energy degradation of the projectile parton, neglected in the CGC, through either elastic scattering or induced gluon brehmstralung would be larger in a nucleus than in proton on account of the stronger color fields of the former, resulting in the relative suppression observed in data. A successful description of forward ratios based on the energy loss calculation in \cite{Kopeliovich:2005ym} is shown in
\fig{ratios}b. Calculating forward di-hadron correlations in both frameworks could help pin down which is the correct picture. In the following, we show that the CGC calculations predicts correctly the azimuthal distribution.

\subsection{Di-hadron azimuthal correlations}

The kinematic range for forward particle detection at RHIC is such that $x_p\!\sim\!0.4$ and
$x_A\!\sim\!10^{-3}.$ Therefore the dominant partonic subprocess is initiated by valence quarks in the proton and, at lowest order in $\alpha_s,$ the $pA\!\to\!h_1h_2X$ double-inclusive cross-section is obtained from the $qA\to qgX$ cross-section, the valence quark density in the proton $f_{q/p}$, and the appropriate hadron fragmentation functions $D_{h/q}$ and $D_{h/g}$:
\begin{eqnarray}
dN^{pA\to h_1 h_2 X}&=&\int_{x_1}^1 dz_1 \int_{x_2}^1 dz_2 \int_{\frac{x_1}{z_1}+\frac{x_2}{z_2}}^1 dx\
f_{q/p}(x,\mu^2)\nonumber\\&&\times\left[dN^{qA\to qgX}\left(xP,\frac{p_1}{z_1},\frac{p_2}{z_2}\right)
D_{h_1/q}(z_1,\mu^2)D_{h_2/g}(z_2,\mu^2)+\right.\nonumber\\&&\left.
dN^{qA\to qgX}\left(xP,\frac{p_2}{z_2},\frac{p_1}{z_1}\right)D_{h_1/g}(z_1,\mu^2)D_{h_2/q}(z_2,\mu^2)\right]\ .
\label{collfact}
\end{eqnarray}
We shall use CTEQ6 NLO quark distributions \cite{Pumplin:2002vw} and KKP NLO fragmentation functions
\cite{Kniehl:2000fe}. The factorization and fragmentation scales are both chosen equal to the transverse momentum of the leading hadron, which we choose to denote hadron 1, $\mu=|p_{1\perp}|.$ We have assumed independent fragmentation of the two final-state hadrons, therefore \eq{collfact} cannot be used to compute the near-side peak around $\Delta\Phi=0$. Doing so would require the use of poorly-known di-hadron fragmentation functions, rather we will focus on the away-side peak around $\Delta\Phi=\pi$ where saturation effects are important.

For the proton, one has $x_p<x<1$, and if $x_p$ would be smaller (this will be the case at the LHC), the gluon initiated processes $gA\to q\bar{q}X$ and $gA\to ggX$ should also be included in \eq{collfact}. For the nucleus, we shall see that the parton momentum fraction varies between $x_A$ and $e^{-2y_1}+e^{-2y_2}$. 
Therefore with large enough rapidities, only the small$-x$ part of the nuclear wave function is relevant when calculating the $qA\to qgX$ cross section, and that cross section cannot be factorized further:
$dN^{qA\to qgX}\neq f_{g/A}\otimes dN^{qg\to qgX}$. Indeed, when probing the saturation regime,
$dN^{qA\to qgX}$ is expected to be a non-linear function of the nuclear gluon distribution, which is itself, through evolution, a non-linear function of the gluon distribution at higher $x$.

Using the CGC approach to describe the small$-x$ part of the nucleus wave function, the $qA\to qgX$ cross section was calculated in \cite{JalilianMarian:2004da,Nikolaev:2005dd,Baier:2005dv,Marquet:2007vb}. It was found that the nucleus cannot be described by only the single-gluon distribution, a direct consequence of the fact that small-x gluons in the nuclear wave function behave coherently, and not individually.
The $qA\!\to\!qgX$ cross section is instead expressed in terms of correlators of Wilson lines (which account for multiple scatterings), with up to a six-point correlator averaged over the CGC wave function.
At the moment, it is not known how to practically evaluate the six-point function. In the large-$N_c$ limit,
it factorizes into a dipole scattering amplitude times a trace of four Wilson lines, and the latter can in principle be obtained by solving an evolution equation written down in \cite{JalilianMarian:2004da}. However, this implies a significant amount of numerical work and also requires to introduce an unknown initial condition.

Rather we shall follow the approach of \cite{Marquet:2007vb}, and use an approximation that allows to express the six-point function in terms of the two-point function \eq{ugdfund}. The resulting cross section for the inclusive production of the quark-gluon system in the scattering of a quark with momentum $xP^+$ off the nucleus $A$ reads \cite{Marquet:2007vb}:
\begin{eqnarray}
\frac{dN^{qA\to qgX}}{d^3kd^3q}&=&
\frac{\alpha_S C_F}{4\pi^2}\ \delta(xP^+\!-\!k^+\!-\!q^+)\ F(\tilde{x}_A,\Delta)
\nonumber\\&&\times\sum_{\lambda\alpha\beta}
\left|I^{\lambda}_{\alpha\beta}(z,k_\perp\!-\!\Delta;{\tilde{x}_A})\!-\!
\psi^{\lambda}_{\alpha\beta}(z,k_\perp\!-\!z\Delta)\right|^2\ ,
\label{cs}\end{eqnarray}
where $q$ and $k$ are the momenta the quark and gluon respectively, and with $\Delta=k_\perp+q_\perp$ and
$z=k^+/xP^+$. In this formula, $\tilde{x}_A$ denotes the longitudinal momentum fraction of the gluon in the nucleus, and $\tilde{x}_A=x_1\ e^{-2y_1}/z_1+x_2\ e^{-2y_2}/z_2>x_A$ when the cross section \eq{cs} is plugged into \eq{collfact}.

\begin{figure}[t]
\begin{center}
\includegraphics[height=5cm]{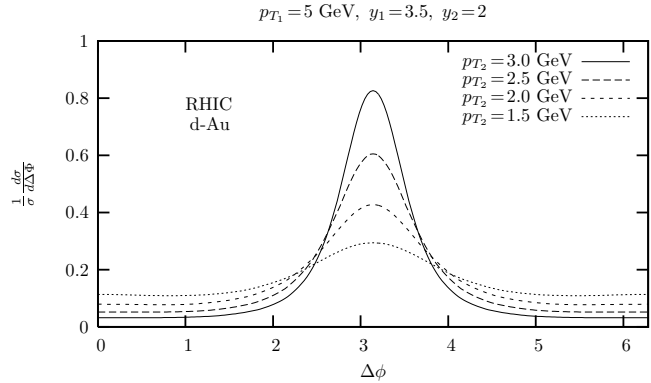}
\end{center}
\caption{The $\Delta\phi$ distribution $(1/\sigma)d\sigma/d\Delta\phi$ computed in \cite{Marquet:2007vb} at fixed $y_1=3.5$, $y_2=2$ and $|p_{1\perp}|=5$ GeV. The azimuthal correlation is suppressed as $|p_{2\perp}|$ decreases, this is due to the onset of non-linearities in the nuclear wave function.}
\label{pred}
\end{figure}

The second line of \eq{cs} features a $k_T$-factorization breaking term:
\begin{equation}
I^{\lambda}_{\alpha\beta}(z,k_\perp;x)
=\int d^2q_\perp \psi^{\lambda}_{\alpha\beta}(z,q_\perp) F(x,k_\perp\!-\!q_\perp)\ ,
\label{split}\end{equation}
and where $\psi^{\lambda}_{\alpha\beta}$ is the well-known amplitude for $q\!\to\!qg$ splitting ($\lambda,$ $\alpha$ and $\beta$ are polarization and helicity indices). While no additionnal information than the two-point function is needed to compute \eq{cs}, since higher-point correlators needed in principle have been expressed in terms of $F(x,k_\perp)$, the cross section is still a non-linear function of that gluon distribution, invalidating $k_T-$factorization. The rather simple form of the $k_T-$factorization breaking term is due to the use of a Gaussian CGC color source distribution, and to the large$-N_c$ limit. Finally, the factor $\delta(xP^+\!-\!k^+\!-\!q^+)$ in \eq{cs} is due to the fact that the eikonal approximation, valid at high energies, is used to compute the $qA\to qgX$ cross section. The delta function is a manifestation of the fact that in a high-energy hadronic collision, the momentum transfer is mainly transverse.

Before comparing our predictions to the recent RHIC data, we first display in \fig{pred} the predictions
for the $\Delta\phi$ distributions made in \cite{Marquet:2007vb} when only the leading-order BK evolution was known. The different curves are obtained for fixed $y_1$, $y_2$ and $|p_{1\perp}|$, while $|p_{2\perp}|$ is varied. Although these results are only qualitative since for instance parton fragmentation was not included, the suppression of the $\Delta\phi$ distribution as $|p_{2\perp}|$ gets closer to the saturation scale was predicted. 

Then, the derivation of the rcBK equation made robust quantitative calculations possible. Indeed, after the parameters of the theory ($x_0$ and $\bar{Q}_{s0}$ for both the proton and the gold nucleus) were contrained by single-inclusive forward spectra, parameter-free predictions for the coincidence probability $CP(\Delta\phi)$ could be made \cite{Albacete:2010pg}. In \eq{kinint}, when computing $N_{pair}$ from \eq{collfact} and
$N_{trig}$ from \eq{hyb} for d+Au and p+p collisions, the integration bounds for the rapidities are $2.4<y<4$, in order to compare with RHIC data \cite{Braidot:2010zh}. In addition, such a restriction does insure that only small-momentum partons are relevant in the nucleus wave function, as is assumed in our calculation. For the trigger (leading) particle $|p_{1\perp}|>2$ GeV and for the associated (sub-leading) hadron
$1\ \mbox{GeV}<|p_{2\perp}|<|p_{1\perp}|.$

\begin{figure}[t]
\begin{center}
\includegraphics[height=5.9cm]{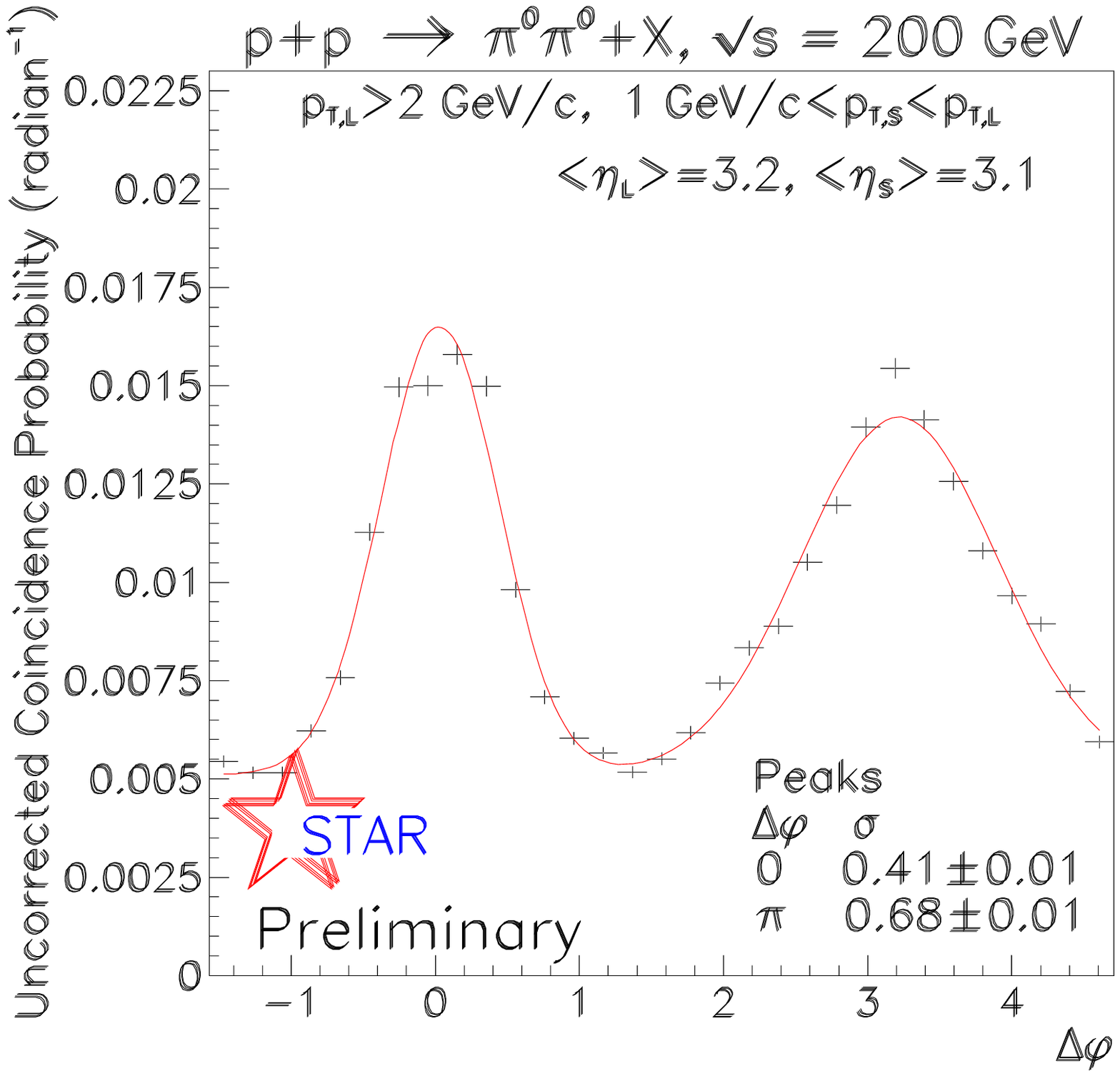}
\hfill
\includegraphics[height=5.9cm]{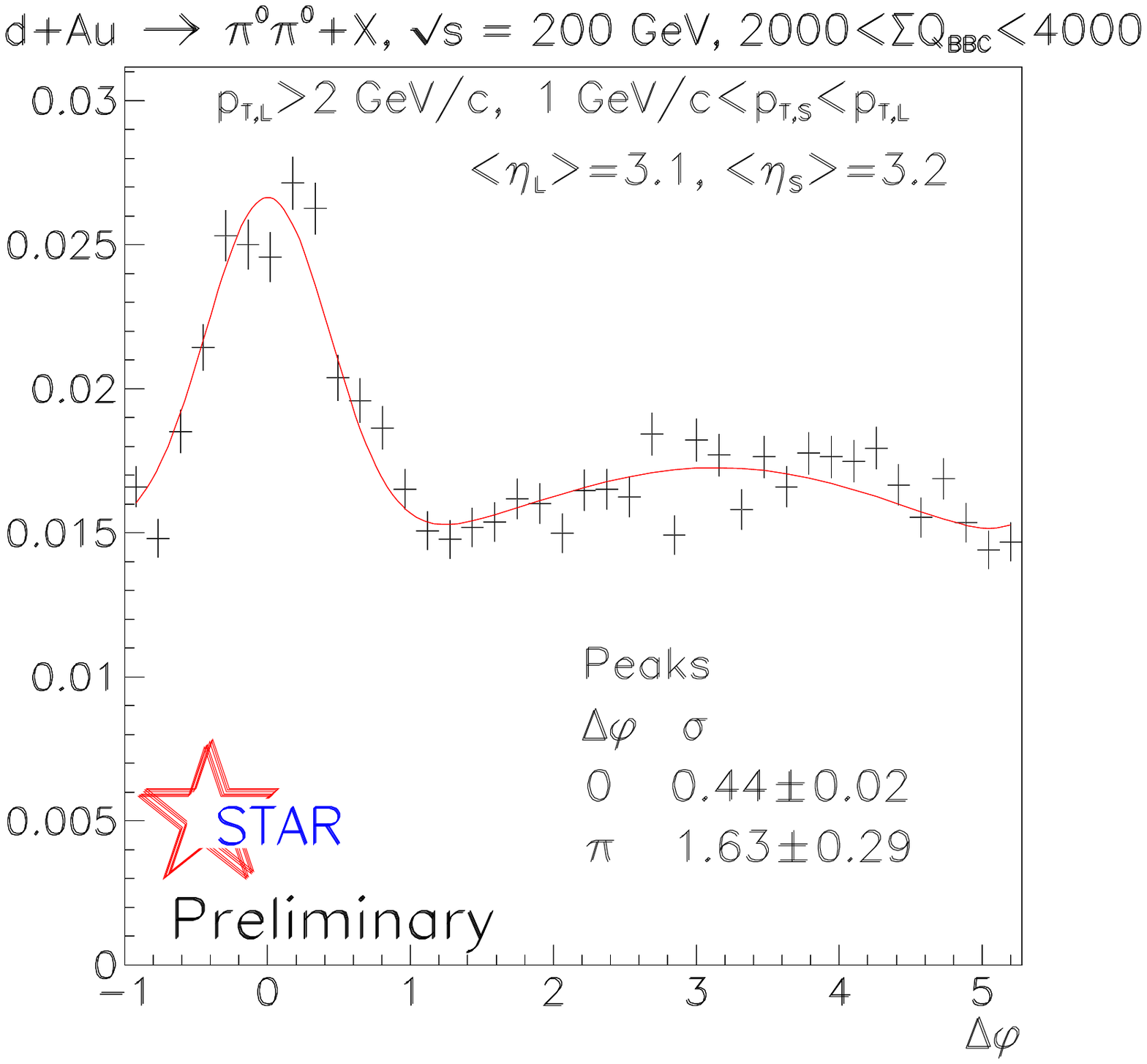}
\end{center}
\caption{The coincidence probability as a function of $\Delta\phi$ for p+p (left) and central d+Au (right) collisions. These preliminary data from \cite{Braidot:2010zh} show a striking nuclear modification of
di-hadron azimuthal correlations. The away-side peak, corresponding to hadrons emitted back-to-back, is prominent in p+p collisions but is absent in the central d+Au case. Such production of monojets was
anticipated in the CGC as a signal of parton saturation.}
\label{starCP}
\end{figure}

\begin{figure}[t]
\begin{center}
\includegraphics[height=4.5cm]{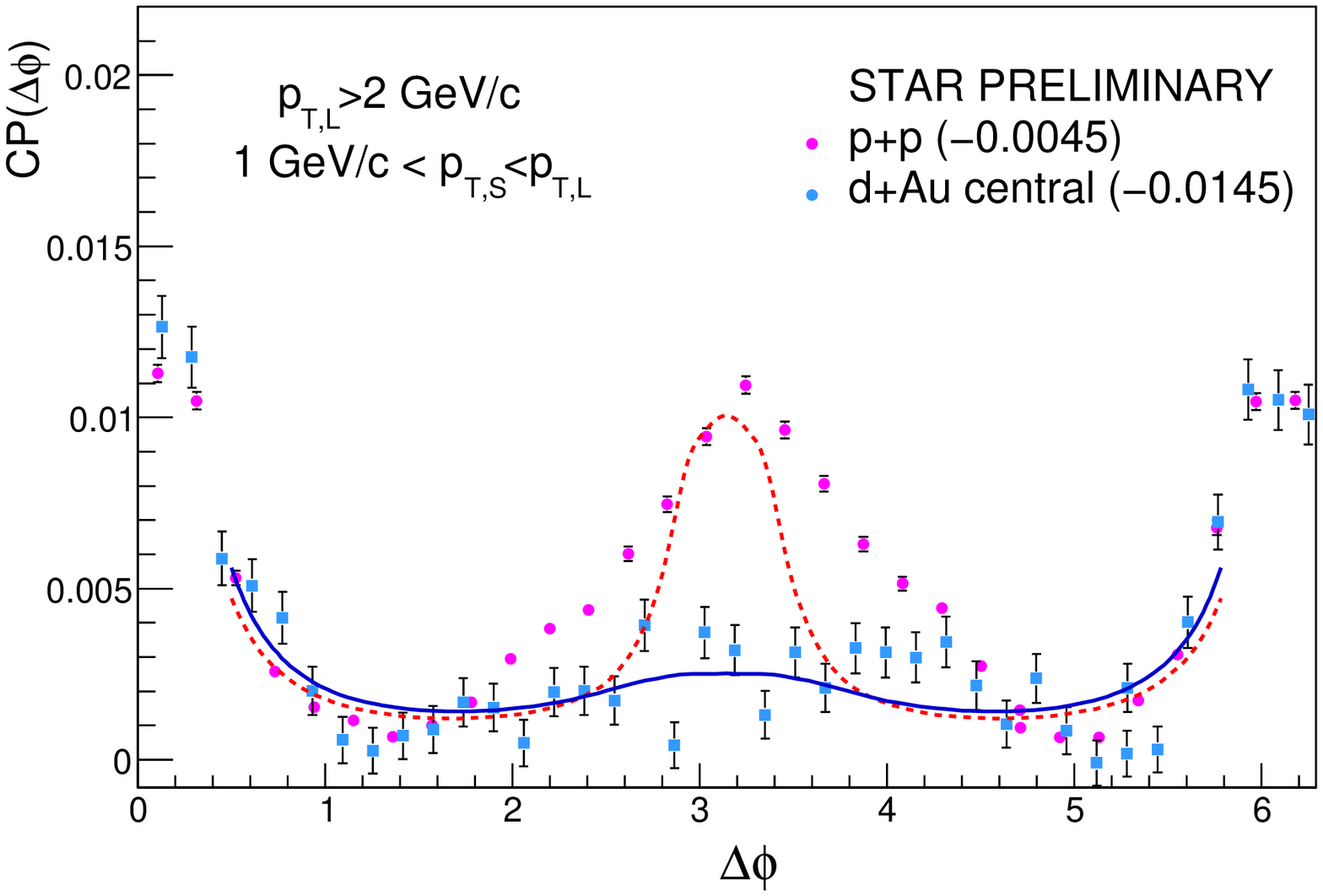}
\hfill
\includegraphics[height=4.5cm]{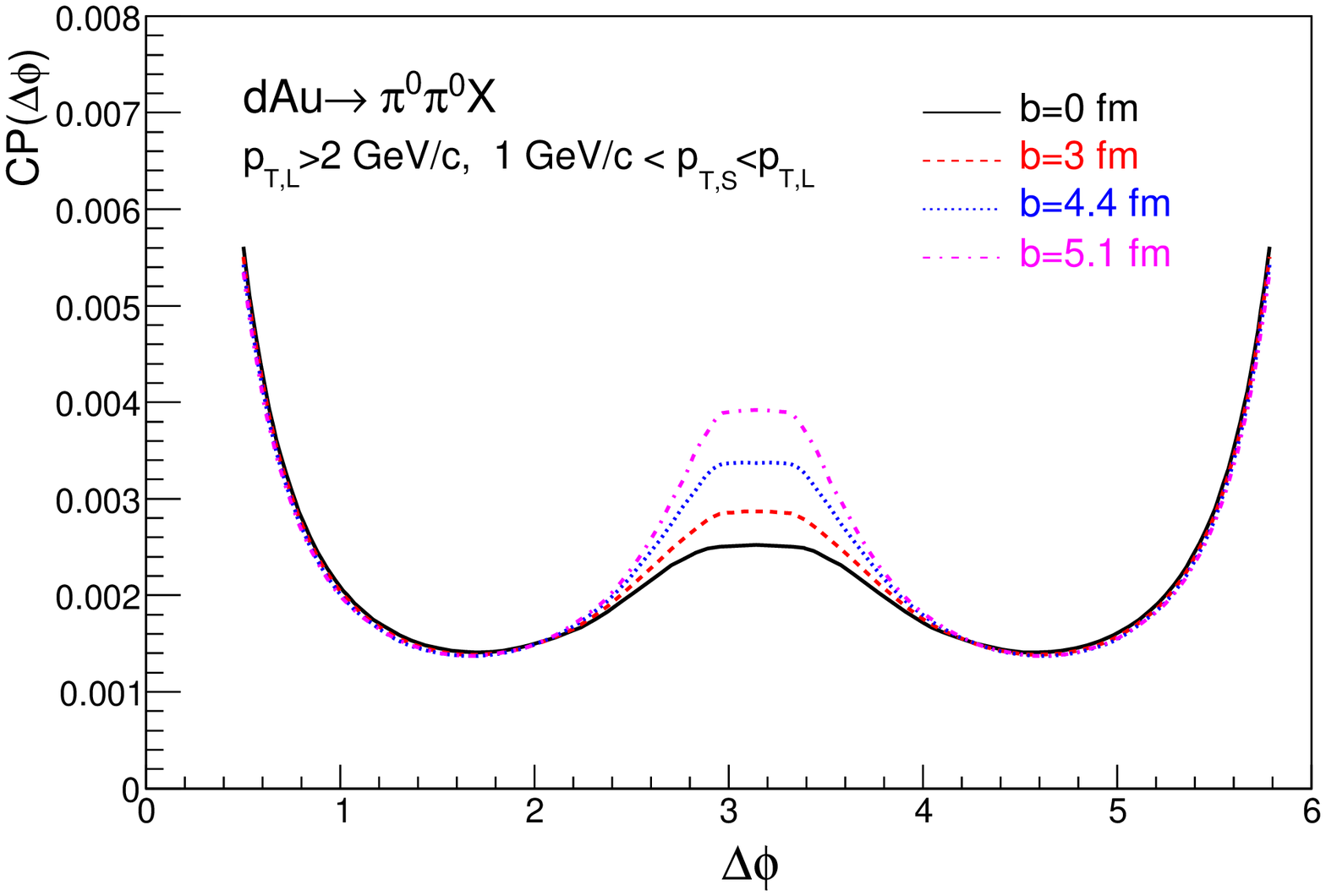}
\end{center}
\caption{The coincidence probability as a function of $\Delta\phi$. Left: for p+p and central d+Au collisions, preliminary data are compared with CGC predictions; the away-side peak in p+p is qualitatively described by the CGC calculation while its disapearence in central d+Au is quantitatively consistent with the prediction. Right: CGC predictions for different centralities of the d+Au collision; the near-side peak is independent of the centrality, while the away-side peak reappears as collisions are more and more peripheral.}
\label{cgcCP}
\end{figure}

The recent data obtained by the STAR collaboration for the coincidence probability obtained with two neutral pions are displayed in \fig{starCP} for both p+p and central d+Au collisions. The nuclear modification of the di-pion azimuthal correlation is quite impressive, the prominent away-side peak in p+p collisions is absent in
central d+Au collisions, in agreement with the behavior predicted in \cite{Marquet:2007vb}.
In \fig{cgcCP}a, these data are compared with the rcBK calculations of \cite{Albacete:2010pg}. As mentioned before, the complete near-side peak is not computed, as \eq{collfact} does not apply around $\Delta\phi=0$.

We see that the disapearence of the away-side peak in central d+Au collisions, compared to p+p collisions, is quantitatively consistent with the CGC calculations. The latter are only robust for the d+Au case, but the extrapolation to the p+p case is displayed in order to show that it is qualitatively consistent with the presence of the away-side peak in p+p, and also with the fact that the near-side peak is identical in the two cases, and is not sensitive to saturation physics. Note that since uncorrelated background has not been extracted from the data, the overall normalization of the data points has been adjusted by subtracting a constant shift, as indicated on the figure.

To deal with the centrality dependence, we have identified the centrality averaged initial saturation scale
$\bar{Q}^2_{s0}$, extracted from minimum-bias single-inclusive hadron production data, with the value of
$Q^2_{s0}$ at $b=5.47$ fm, and used the Woods-Saxon distribution $T_A(b)$ to calculate the saturation scale at other centralities: 
\begin{equation}
Q^2_{s0}(b)=\frac{\bar{Q}_{s0}^2\ T_A(b)}{T_A(5.47\ \mbox{fm})}\ ,\quad\bar{Q}^2_{s0}=0.4\ \mbox{GeV}^2\ .
\end{equation}
The value used in \fig{cgcCP}a in the central d+Au case is $Q^2_{s0}(0)\simeq 0.6$ GeV$^2$ at $x_0=0.02$. The corresponding saturation scale felt by gluons is about $1.2$ GeV$^2$ and of course it gets bigger with decreasing $x$.

In \fig{cgcCP}b, we show the centrality dependence of the coincidence probability. Although it is difficult to trust our formalism all the way to peripheral collisions, we predict that the near-side peak does not change with centrality, and that the away-side peak reappears for less central collisions. This is consistent with the fact that peripheral d+Au collisions are p+p collisions. The fact that the away-side peak disappears from peripheral to central collisions shows that indeed this effect is correlated with the nuclear density.

Moreover dihadron correlations at mid-rapidity, which are sensitive to larger values of $x_A$, feature an away-side peak whatever the centrality. The fact that for central collisions the away-side peak disappears from central to forward rapidities also indicates that the effect is correlated with the nuclear gluon density. In a similar way, we predict that for higher transverse momenta, the away-side peak will reappear.

We are not aware of any descriptions of this phenomena that does not invoke saturation effects. We note that apart from our CGC calculation, a successful description based on the KLN saturation model was also recently proposed \cite{Tuchin:2009nf}. There, although different assumptions are used, the existence of the saturation scale is the crucial ingredient to successfully reproduce the data. While more differential measurements of the coincidence probability, as a function of transverse momentum or rapidity, will provide further quantitative tests of our CGC predictions, the piece of evidence we have discussed in this work strongly indicates that we have observed manifestations of the saturation regime of QCD at RHIC.

\section{Predictions for the LHC}

\begin{figure}[t]
\begin{center}
\includegraphics[height=6.2cm]{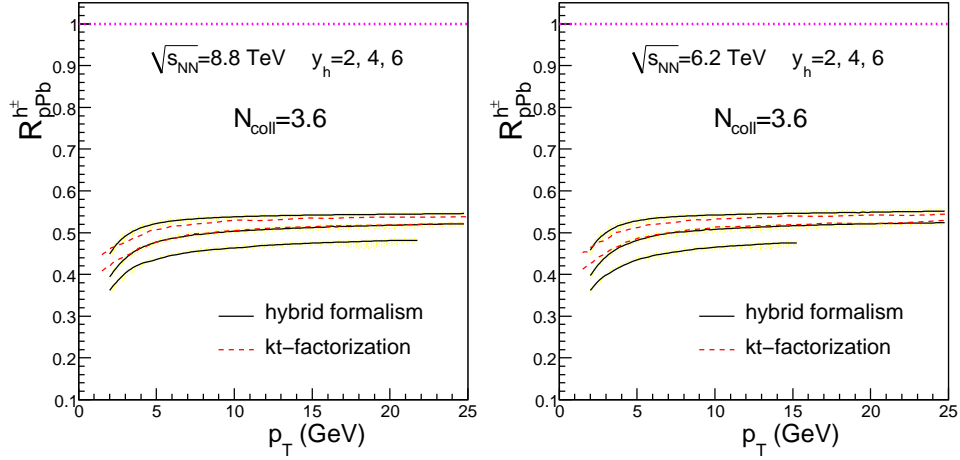}
\caption{CGC predictions for the nuclear modification factor in p+Pb collisions at two LHC energies and rapidities $y=2,4$ and 6.}
\label{LHC}
\end{center}
\end{figure}

The huge leap forward in collision energy reached at the LHC allows for an exploration of small-$x$ effects already at mid-rapidity. There, both the target and projectile are probed at small values of $x$, and energy loss effects associated to large-$x_F$ effects are expected to be small. In \fig{LHC} we present our CGC predictions for the nuclear modification factor for negative charged hadrons in p+Pb collisions at two LHC energies. Our curves correspond to rapidities $y=2$ and larger. Technical difficulties related to the intrinsic asymmetry of the hybrid formalism used for particle production prevent us from calculating the ratios at mid-rapidity. However, the smooth rapidity dependence suggests that a large suppression $\sim 0.6$ is also expected at mid-rapidity in the LHC. It should be taken into account that the normalization taken to produce the curves in \fig{LHC} was $N_{coll}=3.6$.

We also compare the $y=2$ and 4 curves with predictions obtained with the $k_T$-factorization formalism, in order to check the validity of that approach, and especially to test up to what value of $y$ it can be used. The $k_T$-factorization formula (see \cite{Albacete:2010bs}) is valid when the dominant contributions to the cross section come from small values of x, for both the projectile ($x_p\ll1$) and the target ($x_A\ll1$). For instance, it only includes gluonic degrees of freedom. This approach is clearly insufficient at very forward rapidities or large $p_\perp$, where valence quarks of the projectile are important ($x_p\to 1$).
However, as can be seen in \fig{LHC}, both formalisms give comparable results, as the lines from
$k_T$-factorization overlap with the uncertainty bands spanned by the results from the hybrid formalism. This seems to identify a kinematical window where both approximations are valid.

\begin{thebibliography}{38}
\expandafter\ifx\csname natexlab\endcsname\relax\def\natexlab#1{#1}\fi
\providecommand{\bibinfo}[2]{#2}
\ifx\xfnm\relax \def\xfnm[#1]{\unskip,\space#1}\fi
\bibitem[{Gelis et~al.(2010)Gelis, Iancu, Jalilian-Marian, and
  Venugopalan}]{Gelis:2010nm}
\bibinfo{author}{F.~Gelis}, \bibinfo{author}{E.~Iancu},
  \bibinfo{author}{J.~Jalilian-Marian}, \bibinfo{author}{R.~Venugopalan},
\newblock \bibinfo{title}{{The Color Glass Condensate}}
  (\bibinfo{year}{2010}).
\bibitem[{Weigert(2005)}]{Weigert:2005us}
\bibinfo{author}{H.~Weigert},
\newblock \bibinfo{title}{Evolution at small x: The color glass condensate},
\newblock \bibinfo{journal}{Prog. Part. Nucl. Phys.} \bibinfo{volume}{55}
  (\bibinfo{year}{2005}) \bibinfo{pages}{461--565}.
\bibitem[{Arsene et~al.(2004)}]{Arsene:2004ux}
\bibinfo{author}{I.~Arsene}, et~al.,
\newblock \bibinfo{title}{{On the evolution of the nuclear modification factors
  with rapidity and centrality in d + Au collisions at s(NN)**(1/2) =
  200-GeV}},
\newblock \bibinfo{journal}{Phys. Rev. Lett.} \bibinfo{volume}{93}
  (\bibinfo{year}{2004}) \bibinfo{pages}{242303}.
\bibitem[{Adams et~al.(2006)}]{Adams:2006uz}
\bibinfo{author}{J.~Adams}, et~al.,
\newblock \bibinfo{title}{{Forward neutral pion production in p+p and d+Au
  collisions at s(NN)**(1/2) = 200-GeV}},
\newblock \bibinfo{journal}{Phys. Rev. Lett.} \bibinfo{volume}{97}
  (\bibinfo{year}{2006}) \bibinfo{pages}{152302}.
\bibitem[{Braidot and collaboration(2010)}]{Braidot:2010zh}
\bibinfo{author}{E.~Braidot}, \bibinfo{author}{f.~t.~S. collaboration},
\newblock \bibinfo{title}{{Suppression of Forward Pion Correlations in d+Au
  Interactions at STAR}}  (\bibinfo{year}{2010}).
\bibitem[{Albacete and Marquet(2010{\natexlab{a}})}]{Albacete:2010bs}
\bibinfo{author}{J.~L. Albacete}, \bibinfo{author}{C.~Marquet},
\newblock \bibinfo{title}{{Single Inclusive Hadron Production at RHIC and the
  LHC from the Color Glass Condensate}},
\newblock \bibinfo{journal}{Phys. Lett.} \bibinfo{volume}{B687}
  (\bibinfo{year}{2010}{\natexlab{a}}) \bibinfo{pages}{174--179}.
\bibitem[{Albacete and Marquet(2010{\natexlab{b}})}]{Albacete:2010pg}
\bibinfo{author}{J.~L. Albacete}, \bibinfo{author}{C.~Marquet},
\newblock \bibinfo{title}{{Azimuthal correlations of forward di-hadrons in d+Au
  collisions at RHIC in the Color Glass Condensate}}
  (\bibinfo{year}{2010}{\natexlab{b}}).
\bibitem[{Kharzeev et~al.(2003)Kharzeev, Kovchegov, and
  Tuchin}]{Kharzeev:2003wz}
\bibinfo{author}{D.~Kharzeev}, \bibinfo{author}{Y.~V. Kovchegov},
  \bibinfo{author}{K.~Tuchin},
\newblock \bibinfo{title}{Cronin effect and high-p(t) suppression in p a
  collisions},
\newblock \bibinfo{journal}{Phys. Rev.} \bibinfo{volume}{D68}
  (\bibinfo{year}{2003}) \bibinfo{pages}{094013}.
\bibitem[{Albacete et~al.(2004)Albacete, Armesto, Kovner, Salgado, and
  Wiedemann}]{Albacete:2003iq}
\bibinfo{author}{J.~L. Albacete}, \bibinfo{author}{N.~Armesto},
  \bibinfo{author}{A.~Kovner}, \bibinfo{author}{C.~A. Salgado},
  \bibinfo{author}{U.~A. Wiedemann},
\newblock \bibinfo{title}{{Energy dependence of the Cronin effect from
  non-linear QCD evolution}},
\newblock \bibinfo{journal}{Phys. Rev. Lett.} \bibinfo{volume}{92}
  (\bibinfo{year}{2004}) \bibinfo{pages}{082001}.
\bibitem[{Marquet(2007)}]{Marquet:2007vb}
\bibinfo{author}{C.~Marquet},
\newblock \bibinfo{title}{{Forward inclusive dijet production and azimuthal
  correlations in pA collisions}},
\newblock \bibinfo{journal}{Nucl. Phys.} \bibinfo{volume}{A796}
  (\bibinfo{year}{2007}) \bibinfo{pages}{41--60}.
\bibitem[{Eskola et~al.(2009)Eskola, Paukkunen, and Salgado}]{Eskola:2009uj}
\bibinfo{author}{K.~J. Eskola}, \bibinfo{author}{H.~Paukkunen},
  \bibinfo{author}{C.~A. Salgado},
\newblock \bibinfo{title}{{EPS09 - a New Generation of NLO and LO Nuclear
  Parton Distribution Functions}},
\newblock \bibinfo{journal}{JHEP} \bibinfo{volume}{04} (\bibinfo{year}{2009})
  \bibinfo{pages}{065}.
\bibitem[{Accardi and Gyulassy(2004)}]{Accardi:2003jh}
\bibinfo{author}{A.~Accardi}, \bibinfo{author}{M.~Gyulassy},
\newblock \bibinfo{title}{{Cronin effect vs. geometrical shadowing in d + Au
  collisions at RHIC}},
\newblock \bibinfo{journal}{Phys. Lett.} \bibinfo{volume}{B586}
  (\bibinfo{year}{2004}) \bibinfo{pages}{244--253}.
\bibitem[{Kharzeev et~al.(2004)Kharzeev, Kovchegov, and
  Tuchin}]{Kharzeev:2004yx}
\bibinfo{author}{D.~Kharzeev}, \bibinfo{author}{Y.~V. Kovchegov},
  \bibinfo{author}{K.~Tuchin},
\newblock \bibinfo{title}{{Nuclear modification factor in d + Au collisions:
  Onset of suppression in the color glass condensate}},
\newblock \bibinfo{journal}{Phys. Lett.} \bibinfo{volume}{B599}
  (\bibinfo{year}{2004}) \bibinfo{pages}{23--31}.
\bibitem[{Qiu and Vitev(2006)}]{Qiu:2004da}
\bibinfo{author}{J.-w. Qiu}, \bibinfo{author}{I.~Vitev},
\newblock \bibinfo{title}{{Coherent QCD multiple scattering in proton nucleus
  collisions}},
\newblock \bibinfo{journal}{Phys. Lett.} \bibinfo{volume}{B632}
  (\bibinfo{year}{2006}) \bibinfo{pages}{507--511}.
\bibitem[{Kharzeev et~al.(2005)Kharzeev, Levin, and McLerran}]{Kharzeev:2004bw}
\bibinfo{author}{D.~Kharzeev}, \bibinfo{author}{E.~Levin},
  \bibinfo{author}{L.~McLerran},
\newblock \bibinfo{title}{{Jet azimuthal correlations and parton saturation in
  the color glass condensate}},
\newblock \bibinfo{journal}{Nucl. Phys.} \bibinfo{volume}{A748}
  (\bibinfo{year}{2005}) \bibinfo{pages}{627--640}.
\bibitem[{Adler et~al.(2006)}]{Adler:2006hi}
\bibinfo{author}{S.~S. Adler}, et~al.,
\newblock \bibinfo{title}{{Azimuthal angle correlations for rapidity separated
  hadron pairs in d + Au collisions at s(NN)**(1/2) = 200-GeV}},
\newblock \bibinfo{journal}{Phys. Rev. Lett.} \bibinfo{volume}{96}
  (\bibinfo{year}{2006}) \bibinfo{pages}{222301}.
\bibitem[{Meredith(2009)}]{Meredith:2009fp}
\bibinfo{author}{B.~Meredith},
\newblock \bibinfo{title}{{Probing High Parton Densities at Low-$x$ in d+Au
  Collisions at PHENIX Using the New Forward and Backward Muon Piston
  Calorimeters}},
\newblock \bibinfo{journal}{Nucl. Phys.} \bibinfo{volume}{A830}
  (\bibinfo{year}{2009}) \bibinfo{pages}{595c--598c}.
\bibitem[{Balitsky(1996)}]{Balitsky:1995ub}
\bibinfo{author}{I.~Balitsky},
\newblock \bibinfo{title}{{Operator expansion for high-energy scattering}},
\newblock \bibinfo{journal}{Nucl. Phys.} \bibinfo{volume}{B463}
  (\bibinfo{year}{1996}) \bibinfo{pages}{99--160}.
\bibitem[{Kovchegov(1999)}]{Kovchegov:1999yj}
\bibinfo{author}{Y.~V. Kovchegov},
\newblock \bibinfo{title}{Small-x {$F_2$} structure function of a nucleus
  including multiple pomeron exchanges},
\newblock \bibinfo{journal}{Phys. Rev.} \bibinfo{volume}{D60}
  (\bibinfo{year}{1999}) \bibinfo{pages}{034008}.
\bibitem[{Balitsky(2007)}]{Balitsky:2006wa}
\bibinfo{author}{I.~I. Balitsky},
\newblock \bibinfo{title}{{Quark Contribution to the Small-$x$ Evolution of
  Color Dipole}},
\newblock \bibinfo{journal}{Phys. Rev. D} \bibinfo{volume}{75}
  (\bibinfo{year}{2007}) \bibinfo{pages}{014001}.
\bibitem[{Kovchegov and Weigert(2007)}]{Kovchegov:2006vj}
\bibinfo{author}{Y.~Kovchegov}, \bibinfo{author}{H.~Weigert},
\newblock \bibinfo{title}{{Triumvirate of Running Couplings in Small-$x$
  Evolution}},
\newblock \bibinfo{journal}{Nucl. Phys. {\bf A}} \bibinfo{volume}{784}
  (\bibinfo{year}{2007}) \bibinfo{pages}{188--226}.
\bibitem[{Albacete et~al.(2009)Albacete, Armesto, Milhano, and
  Salgado}]{Albacete:2009fh}
\bibinfo{author}{J.~L. Albacete}, \bibinfo{author}{N.~Armesto},
  \bibinfo{author}{J.~G. Milhano}, \bibinfo{author}{C.~A. Salgado},
\newblock \bibinfo{title}{{Non-linear QCD meets data: A global analysis of
  lepton- proton scattering with running coupling BK evolution}},
\newblock \bibinfo{journal}{Phys. Rev.} \bibinfo{volume}{D80}
  (\bibinfo{year}{2009}) \bibinfo{pages}{034031}.
\bibitem[{Albacete(2007)}]{Albacete:2007sm}
\bibinfo{author}{J.~L. Albacete},
\newblock \bibinfo{title}{{Particle multiplicities in Lead-Lead collisions at
  the LHC from non-linear evolution with running coupling}},
\newblock \bibinfo{journal}{Phys. Rev. Lett.} \bibinfo{volume}{99}
  (\bibinfo{year}{2007}) \bibinfo{pages}{262301}.
\bibitem[{Albacete and Kovchegov(2007)}]{Albacete:2007yr}
\bibinfo{author}{J.~L. Albacete}, \bibinfo{author}{Y.~V. Kovchegov},
\newblock \bibinfo{title}{Solving high energy evolution equation including
  running coupling corrections},
\newblock \bibinfo{journal}{Phys. Rev.} \bibinfo{volume}{D75}
  (\bibinfo{year}{2007}) \bibinfo{pages}{125021}.
\bibitem[{Dumitru et~al.(2006)Dumitru, Hayashigaki, and
  Jalilian-Marian}]{Dumitru:2005gt}
\bibinfo{author}{A.~Dumitru}, \bibinfo{author}{A.~Hayashigaki},
  \bibinfo{author}{J.~Jalilian-Marian},
\newblock \bibinfo{title}{{The color glass condensate and hadron production in
  the forward region}},
\newblock \bibinfo{journal}{Nucl. Phys.} \bibinfo{volume}{A765}
  (\bibinfo{year}{2006}) \bibinfo{pages}{464--482}.
\bibitem[{Kovner and Wiedemann(2001)}]{Kovner:2001vi}
\bibinfo{author}{A.~Kovner}, \bibinfo{author}{U.~A. Wiedemann},
\newblock \bibinfo{title}{{Eikonal evolution and gluon radiation}},
\newblock \bibinfo{journal}{Phys. Rev.} \bibinfo{volume}{D64}
  (\bibinfo{year}{2001}) \bibinfo{pages}{114002}.
\bibitem[{Kovchegov and Tuchin(2002)}]{Kovchegov:2001sc}
\bibinfo{author}{Y.~V. Kovchegov}, \bibinfo{author}{K.~Tuchin},
\newblock \bibinfo{title}{Inclusive gluon production in dis at high parton
  density},
\newblock \bibinfo{journal}{Phys. Rev.} \bibinfo{volume}{D65}
  (\bibinfo{year}{2002}) \bibinfo{pages}{074026}.
\bibitem[{Marquet(2005)}]{Marquet:2004xa}
\bibinfo{author}{C.~Marquet},
\newblock \bibinfo{title}{{A QCD dipole formalism for forward-gluon
  production}},
\newblock \bibinfo{journal}{Nucl. Phys.} \bibinfo{volume}{B705}
  (\bibinfo{year}{2005}) \bibinfo{pages}{319--338}.
\bibitem[{Pumplin et~al.(2002)}]{Pumplin:2002vw}
\bibinfo{author}{J.~Pumplin}, et~al.,
\newblock \bibinfo{title}{{New generation of parton distributions with
  uncertainties from global QCD analysis}},
\newblock \bibinfo{journal}{JHEP} \bibinfo{volume}{07} (\bibinfo{year}{2002})
  \bibinfo{pages}{012}.
\bibitem[{de~Florian et~al.(2007{\natexlab{a}})de~Florian, Sassot, and
  Stratmann}]{deFlorian:2007aj}
\bibinfo{author}{D.~de~Florian}, \bibinfo{author}{R.~Sassot},
  \bibinfo{author}{M.~Stratmann},
\newblock \bibinfo{title}{{Global analysis of fragmentation functions for pions
  and kaons and their uncertainties}},
\newblock \bibinfo{journal}{Phys. Rev.} \bibinfo{volume}{D75}
  (\bibinfo{year}{2007}{\natexlab{a}}) \bibinfo{pages}{114010}.
\bibitem[{de~Florian et~al.(2007{\natexlab{b}})de~Florian, Sassot, and
  Stratmann}]{deFlorian:2007hc}
\bibinfo{author}{D.~de~Florian}, \bibinfo{author}{R.~Sassot},
  \bibinfo{author}{M.~Stratmann},
\newblock \bibinfo{title}{{Global analysis of fragmentation functions for
  protons and charged hadrons}},
\newblock \bibinfo{journal}{Phys. Rev.} \bibinfo{volume}{D76}
  (\bibinfo{year}{2007}{\natexlab{b}}) \bibinfo{pages}{074033}.
\bibitem[{Kopeliovich et~al.(2005)Kopeliovich, Nemchik, Potashnikova, Johnson,
  and Schmidt}]{Kopeliovich:2005ym}
\bibinfo{author}{B.~Z. Kopeliovich}, \bibinfo{author}{J.~Nemchik},
  \bibinfo{author}{I.~K. Potashnikova}, \bibinfo{author}{M.~B. Johnson},
  \bibinfo{author}{I.~Schmidt},
\newblock \bibinfo{title}{{Breakdown of QCD factorization at large Feynman x}},
\newblock \bibinfo{journal}{Phys. Rev.} \bibinfo{volume}{C72}
  (\bibinfo{year}{2005}) \bibinfo{pages}{054606}.
\bibitem[{Frankfurt and Strikman(2007)}]{Frankfurt:2007rn}
\bibinfo{author}{L.~Frankfurt}, \bibinfo{author}{M.~Strikman},
\newblock \bibinfo{title}{{Energy losses in the black disc regime and
  correlation effects in the STAR forward pion production in d Au collisions}},
\newblock \bibinfo{journal}{Phys. Lett.} \bibinfo{volume}{B645}
  (\bibinfo{year}{2007}) \bibinfo{pages}{412--421}.
\bibitem[{Kniehl et~al.(2000)Kniehl, Kramer, and Potter}]{Kniehl:2000fe}
\bibinfo{author}{B.~A. Kniehl}, \bibinfo{author}{G.~Kramer},
  \bibinfo{author}{B.~Potter},
\newblock \bibinfo{title}{{Fragmentation functions for pions, kaons, and
  protons at next-to-leading order}},
\newblock \bibinfo{journal}{Nucl. Phys.} \bibinfo{volume}{B582}
  (\bibinfo{year}{2000}) \bibinfo{pages}{514--536}.
\bibitem[{Jalilian-Marian and Kovchegov(2004)}]{JalilianMarian:2004da}
\bibinfo{author}{J.~Jalilian-Marian}, \bibinfo{author}{Y.~V. Kovchegov},
\newblock \bibinfo{title}{{Inclusive two-gluon and valence quark-gluon
  production in DIS and p A}},
\newblock \bibinfo{journal}{Phys. Rev.} \bibinfo{volume}{D70}
  (\bibinfo{year}{2004}) \bibinfo{pages}{114017}.
\bibitem[{Nikolaev et~al.(2005)Nikolaev, Schafer, Zakharov, and
  Zoller}]{Nikolaev:2005dd}
\bibinfo{author}{N.~N. Nikolaev}, \bibinfo{author}{W.~Schafer},
  \bibinfo{author}{B.~G. Zakharov}, \bibinfo{author}{V.~R. Zoller},
\newblock \bibinfo{title}{{Nonlinear k(T)-factorization for quark-gluon dijet
  production off nuclei}},
\newblock \bibinfo{journal}{Phys. Rev.} \bibinfo{volume}{D72}
  (\bibinfo{year}{2005}) \bibinfo{pages}{034033}.
\bibitem[{Baier et~al.(2005)Baier, Kovner, Nardi, and Wiedemann}]{Baier:2005dv}
\bibinfo{author}{R.~Baier}, \bibinfo{author}{A.~Kovner},
  \bibinfo{author}{M.~Nardi}, \bibinfo{author}{U.~A. Wiedemann},
\newblock \bibinfo{title}{{Particle correlations in saturated QCD matter}},
\newblock \bibinfo{journal}{Phys. Rev.} \bibinfo{volume}{D72}
  (\bibinfo{year}{2005}) \bibinfo{pages}{094013}.
\bibitem[{Tuchin(2010)}]{Tuchin:2009nf}
\bibinfo{author}{K.~Tuchin},
\newblock \bibinfo{title}{{Rapidity and centrality dependence of azimuthal
  correlations in Deuteron-Gold collisions at RHIC}},
\newblock \bibinfo{journal}{Nucl. Phys.} \bibinfo{volume}{A846}
  (\bibinfo{year}{2010}) \bibinfo{pages}{83--94}.

\end{thebibliography}


\end{document}